\documentclass{aa}
\usepackage{txfonts}
\usepackage{graphicx,natbib}

\newcommand{\bl}{BL~Lacertae}

\newcommand{\etal}{et al.}
\def\simlt{\lower.5ex\hbox{\ltsima}}            
\def\simgt{\lower.5ex\hbox{\gtsima}}            

\def\la{~\raise.5ex\hbox{$<$}\kern-.8em\lower 1mm\hbox{$\sim$}~}
\def\ma{~\raise.5ex\hbox{$>$}\kern-.8em\lower 1mm\hbox{$\sim$}~}

\begin{document}
\title{The long-term optical spectral variability of \bl\thanks{Based on
data taken and assembled by the WEBT collaboration and stored in the WEBT
archive at the Osservatorio Astronomico di Torino - INAF.}}

\author{I.~E.~Papadakis\inst{1,2} \and M.~Villata\inst{3}
\and C.~M.~Raiteri\inst{3}}
\offprints{I. E. Papadakis;  e-mail: jhep@physics.uoc.gr}
\institute{Physics Department, University of Crete, P.O. Box 2208,
   710 03 Heraklion, Crete, Greece
\and  IESL, Foundation for Research and Technology, 711 10 Heraklion, Greece
\and INAF, Osservatorio Astronomico di
Torino, via Osservatorio 20, 10025 Pino Torinese (TO), Italy} 
\date{Received ?/ Accepted ?} 
\abstract
{We present the results from a study of the long-term optical spectral
variations of \bl, using the long and well-sampled $B$ and $R$-band light
curves of the Whole Earth Blazar Telescope (WEBT) collaboration, binned on time
intervals of 1 day.}
{To study the relation  between the long-term spectral variations and the
respective flux variations of the source.}
{We use cross-correlation techniques to investigate whether there are any 
delays between the flux variations in different energy bands and between the
flux and spectral variations.}
{The relation between spectral slope and flux (the spectrum gets bluer as the
source flux increases) is well described by a power-law model, although there
is significant scatter around the best-fitting model line. To some extent, this
is due to the spectral evolution of the source (along well-defined loop-like
structures) during low-amplitude events, which are superimposed on the major
optical flares, and evolve on time scales of a few days. The $B$ and $R$-band
variations are well correlated, with no significant, measurable delays larger
than a few days. On the other hand, we find that the spectral variations lead
those in the flux light curves by $\sim 4$ days. Finally, during at least the
largest amplitude  flares, the $B$-band variations appear to evolve faster than
those in the $R$ band.}
{We confirm the ``bluer-when-brighter" mild chromatism of the long-term
variations, and we show that it can be explained if the flux
increases/decreases faster in the $B$ than in the $R$ band. We also report the
discovery of the lag between spectral and flux changes. These two features can
be explained in terms of  Doppler factor variations due to changes in the
viewing angle of a curved and inhomogeneous emitting jet.}
\titlerunning{Optical spectral variability of \bl}
\authorrunning{Papadakis \etal}

\maketitle
   
\section{Introduction}
\smallskip

\bl\ is the prototype of a class of active galactic nuclei (AGNs) known as ``BL
Lac objects", or simply ``BL Lacs". The members of this class show non-thermal
continuum energy spectra, fast and large-amplitude variations from radio up to
$\gamma$-rays, a high degree of linear polarization, and radio jets with
individual components often exhibiting apparent superluminal motion.

The ``Whole Earth Blazar Telescope" (WEBT; {\tt
http://www.to.astro.it/blazars/webt/}) is a large international collaboration
among optical and radio astronomers, established in 1997. It  organizes
monitoring campaigns on selected blazars in order to obtain continuous,
high-temporal-density light curves in the optical and radio bands (usually in
conjunction with observations at other wavelengths like X and $\gamma$-rays;
see e.g. \citealt[]{boe05,rai05,rai06b,ost06,vil06,vil07}). 

\bl\ has been the target of four WEBT monitoring campaigns in the past.  The
first two were rather short. They were organized in 1999, simultaneously with
{\it ASCA} and {\it BeppoSAX} observations \citep{rav02}. The remaining two
were long-term campaigns carried out in the periods May 2000  -- January 2001
\citep{vil02,boe03} and May 2001 -- February 2002 \citep{vil04a,vil04b}. Both
campaigns resulted in long, well-sampled, and high-precision light curves. 

Using data from the first long-term WEBT campaign, \citet{vil02} found that the
$B$ and $R$-band light curves were well correlated with no measurable time
delay. They also found that the flux variations are associated with significant
spectral variations. They interpreted them in the context of a two variability
mechanism model: the first mechanism is essentially achromatic and it is
responsible for the long-term variations (i.e. variations which operate on time
scales of a few days), while the second one causes fast (i.e. shorter than a
day) flares, superposed on the long-term  variations, and introduces spectral
changes in the sense that the spectrum becomes harder (bluer) as the source
brightens. 

Similar results were obtained by \citet{vil04a}, who studied not only the light
curves from the 2001--2002 WEBT campaign, but also composed light curves from
1994 to 2002, in all optical bands, with data taken from the literature as well
as by members of the WEBT. The authors distinguished a ``mildly-chromatic" (in
the sense that the $B-R$ versus $R$ plot has a slope $\sim 0.1$), long-term
variability component, which operates on time scales longer than a few days,
from a ``strongly-chromatic" (in which case the variations trace a 0.4 slope in
the $B-R$ versus $R$ plot), short-term variability mechanism, which operates on
intra-day time scales. 

\citet{vil04a} based their results mainly on the study of the  best-sampled
parts of the light curves, which cover the period between 1997 and 2002. In
this work, we focus our attention on the $B$ and $R$-band  light curves of
\bl\  in the same period.  Our main aim is to better understand the long-term
chromatic behaviour of the source, and to investigate possible reasons for the
observed spectral variations.  

In Sect.\ 2 we present the light curves we use in this work and we quantify
the  ``spectral slope versus flux" relation  on time scales longer than a day.
Sect.\ 3 deals with the cross-correlation analysis between the $B$ and $R$-band
light curves, and between the spectral slope and flux variations. In Sect.\ 4
we present the results from the study of two long, well sampled,
large-amplitude flares in the light curves, while our final conclusions are
discussed in   Sect.\ 5. 

\section{The optical light curves}

\citet{vil04a} presented $UBVRI$ light curves of \bl\ from 1994 up to 2002,
using data from the four WEBT campaigns and the literature. The host galaxy
constant flux contribution was subtracted from the observed light curves in all
bands as explained in Section 5 of Villata \etal, 2002. We chose to use the
\citet{vil04a} $B$ and $R$-band light curves from the period between 1997 and
February 2002, as we believe that these are the ``best" light curves (in terms
of length, number of data points and sampling frequency of observations) among
the currently available optical datasets of this object. The data that we use
in this work are now publicly available in the WEBT archive 
(http://www.to.astro.it/blazars/webt/) and can be requested by sending a
message to the WEBT president. 

In order to minimize any systematic effects that could influence our results we
examined carefully the \citet{vil04a} data and selected our final $B$ and
$R$-band datasets as follows:  a) we kept data with errors not greater than
0.04 and 0.03 mag in the $B$ and  $R$-band light curves, respectively;  b) we
used only those data points which have at least one counterpart in the other
band  within 20 min, either from the same telescope or from the same literature
paper;  c) we also removed three ``unreliable" $B$ data points. 

\begin{figure}
   \centering
   \includegraphics[width=8.5cm]{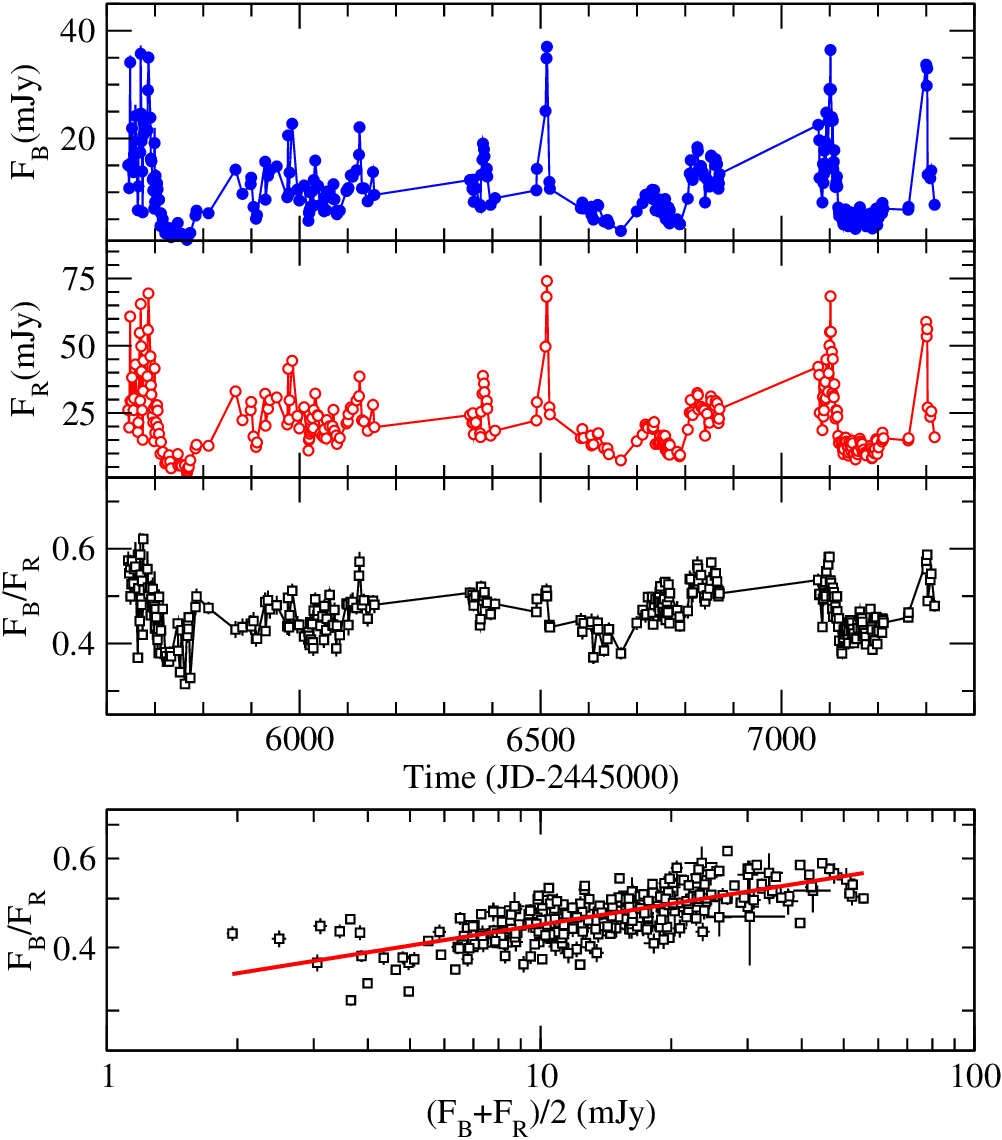}
\caption{First and second panel from top: The 1997--2002 $B$ and $R$ light
curves of \bl\ binned in intervals of size equal to 1 day. Third panel from
top: The $F_B/F_R$ flux density ratio, plotted as a function of time. Bottom
panel: $F_B/F_R$ plotted as a function of $(F_B + F_R)/2$.  The solid line
shows the best fitting power-law model to the data.}
\label{figure:lcs}
\end{figure}

The two upper panels in Fig.~\ref{figure:lcs} show the $B$ and $R$-band light
curves (in flux density units) that we use in this work. Since we are
interested in studying the spectral variations on time scales longer than a
day, the light curves are binned in 1-day long intervals, starting from JD =
2450644. If more than one point is present in the interval, we take their
average flux density and time of observation. We consider the standard
deviation, i.e. $1\sigma$, the average spread of the points around their mean,
as the ``error" on the average flux density in the respective bin.  In the case
there is just one point in a bin we accept its error as it is. Typically, the
error of the points, estimated in this way, in the $B$ and $R$ band 1 day
binned light curves is of the order of $\sim 0.1-1$ mJy. Therefore, although 
we do plot the error bars of the points in  Fig.~\ref{figure:lcs}, most of them
are smaller than the size of the symbols we use.

The source is brighter in the $R$ band. The maximum flux density recorded in
that band is almost twice the respective maximum flux density in the $B$-band
light curve. Apart from this difference, the two light curves look very
similar. The same variations  appear, simultaneously, in both of them. However,
the  amplitude of the observed variations is larger in the $B$-band light
curve. Indeed, the ``fractional variability amplitude", $f_{\rm rms}$ (defined
as the ratio of the standard deviation over the light curve mean, see
\citealt[]{pap03}), of the $B$ and $R$ light curves is $65.3\pm0.2$\% and
$57.1\pm 0.2$\%, respectively.  The errors of $f_{\rm rms}$ account only  for
the measurement errors in the light curve points, and have been estimated
according to the prescription of \citealt{vau03}. Given the very good
correlation between the B  and R band light curves, one could assume that the
$\sim 7$\%  difference in $f_{\rm rms, B}$ and $f_{\rm rms, R}$ B maybe due to
differences in the experimental noise level of the two light curves. For
example, Poisson noise will introduce some scatter in the $f_{\rm rms}$  values
we measure.  However, the errors we provide (following \citealt{vau03}) suggest
that this is not the case. The difference between the $B$ and $R$ band
variability amplitudes is most probably an intrinsic property if the
variability mechanism that opearates in the source.

\subsection{Spectral variations}

The third panel from the top in Fig.~\ref{figure:lcs} shows a plot of the $F_B/F_R$
flux density ratio as a function of time. This ratio can be considered as
representative of the slope of the optical spectrum of the source. Errors on
this ratio have been estimated using the usual ``propagation of errors" recipe
\citep[see e.g.][]{bev69}.  

Since the $B$ and $R$-band light curves do not have the same fractional
variability amplitude, we expect the spectral slope (and hence the  ratio
$F_B/F_R$) to change with time.  Indeed, Fig.~\ref{figure:lcs} shows clearly
that this ratio is highly variable.  The fractional amplitude,  $f_{\rm rms,
R}$, of the $F_B/F_R$ variations is $10.3\pm0.2$\% (where the error accounts
only for the  measurement error of the $F_B/F_R$ points, as explained above).
We conclude that, on time scales longer than 1 day,  the optical flux density
variations of \bl\ are associated with moderate, but significant, spectral
slope variations as well. 

In the bottom panel of Fig.~\ref{figure:lcs} we plot the $F_B/F_R$ ratio as a
function of $(F_B + F_R)/2$, i.e.\ the mean of the $B$ and $R$-band flux
densities,  which can be considered as representative of the flux density in an
intermediate band. We have chosen to use this quantity in the study of the
``spectral slope ($F_B/F_R$) versus brightness" relation, as the average of the
source signal in the two bands should minimize the possibility of introducing
artificial correlations in this relation. Such correlations could be introduced
if we were using either the $B$ or the $R$-band flux densities, individually,
because of the interdependency of the $F_B/F_R$ values  nd these measurements.

The ``spectral slope vs brightness" plot shows clearly that the flux and
spectral variations are well correlated: as the source flux increases
(decreases), the spectrum becomes harder (softer). The solid line in the 
bottom panel of Fig.~\ref{figure:lcs} shows the best fitting power-law (PL)
model to the data: $F_B/F_R \propto [(F_B+F_R)/2]^{0.14 \pm 0.01}$.  Although
it does not provide a statistically accepted fit (we get a $\chi^{2}$ value of 
6313.5 for 308 degrees of freedom when we consider the measurement uncertainty
of each point in the plot), it nevertheless describes rather well the overall
trend of the $F_B/F_R$ increase with increasing flux.  

The statistically poor quality of the fit is due to the fact that  there exists
a significant scatter in the $F_B/F_R$ values for a given flux density.
However, it is of small amplitude (at each flux level, the maximum deviation
around $(F_B/F_R)_{\rm model}$ is less than $\sim \pm 15$\% of $(F_B/F_R)_{\rm
model}$) and does not show any systematic deviations from the best fitting PL
model. This scatter implies some residual, weak spectral variability which is
not correlated with the  flux density level of the source. 

Interestingly, the $F_B/F_R \propto [(F_B+F_R)/2]^{0.14}$ relation would
correspond to a line of slope $\sim 0.14$ in a ``$B-R$ vs  magnitude of $(F_B +
F_R)/2$" plot. This is close to the  ``softer" slope of 0.1 in the ``$B-R$ vs
$R$" plot of \citet{vil04a}.  Hence the flux-density variations that we observe
in the 1-day binned light curves  of Fig.~\ref{figure:lcs} correspond to the
``mildly-chromatic", long-term variations of (\citealt{vil04a}; see also
\citealt{hu06}).

\section{Cross-correlation analysis}

\subsection{Correlation between the $B$ and $R$-band light curves}

\begin{figure}
   \centering
   \includegraphics[width=8.5cm]{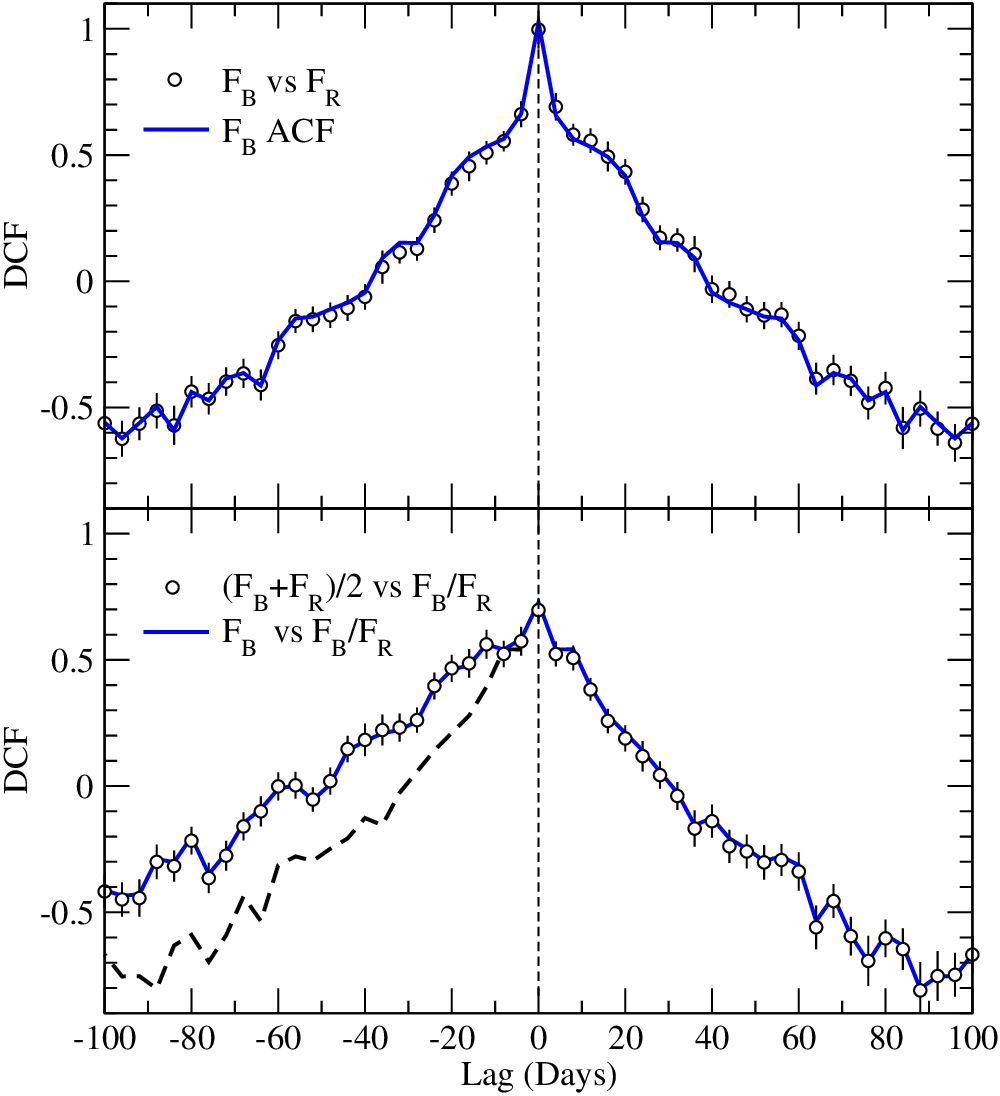}
\caption{Top panel: The $B$-band ACF (solid line) and the $F_B$ vs $F_R$ DCF
(open circles).  Bottom panel: The $(F_B+F_R)/2$ and $F_B$ vs $F_B/F_R$ DCFs
(open circles and solid line, respectively).  The positive-lag $F_B$ vs
$F_B/F_R$ DCF part is reflected about the zero-lag axis  (dashed line) for
comparison reasons (see text for details).} 
\label{figure:ccfs} 
\end{figure}

A possible explanation for the observed spectral variations is the existence of
a delay between the $B$ and $R$-band light curves: if the  flux increases or
decreases with the same rate in both light curves, but the $B$-band variations
lead those in the $R$ band, then the spectrum is ``bluer"/``redder" when the
flux is rising/decaying, respectively. 

In order to investigate this possibility we used the discrete correlation
function (DCF) method of \citet{ede88}. Our results, with a DCF bin size of 4
days,  are plotted in the top panel of Fig.~\ref{figure:ccfs} (open circles). 
We  get similar results when we use a wide range of DCF bin sizes, from 1 up to
10 days. A positive lag of the DCF peak in this plot would imply that the
$B$-band variations lead those in the $R$ band.  In the same panel we also 
plot the $B$-band auto-correlation function (ACF; solid line). The ACF of the
$F_R$ and $(F_B+F_R)/2$ light curves are practically identical to that of the
$F_B$ light curve. 

The $F_B$ vs $F_R$ DCF is similar to the $B$-band ACF and shows a strong 
(DCF$_{\rm max}\approx$ 1)and narrow peak at zero time lag. Strong and narrow
DCF maxima at zero lags can be introduced when systematic errors affect, in the
same way, the data points in both light curves. However, in our case, the
points in the  light curves shown in Fig.~\ref{figure:lcs}  correspond to
observations from many different telescopes, so that the presence of a global
systematic effect is  rather unlikely.  

If we omit the DCF point at zero lag, then the highest DCF value  is $\sim
0.7$. If we now estimate  the centroid lag, $\tau_{\rm cent}$, using all DCF
points with values in excess of 0.7\, DCF$_{\rm max}$ (assuming DCF$_{\rm
max}\sim 0.7$),  we find that $\tau_{\rm cent}=2$ days. This result suggests
that there may exist after all  a small delay between the variations in the $B$
and $R$-band light curves. The positive centroid lag  is due to the fact that
at negative lags (down to $\sim -60$ days) the $F_B$ vs $F_R$ DCF is smaller
than the $F_B$ ACF while the opposite trend, i.e.\ $F_B$ vs $F_R$ DCF being
slightly larger than the $F_B$ ACF, is observed at the positive lags up to
60--70 days. Admittedly, this is a low amplitude effect, and longer light
curves are necessary in order to test its validity.

\subsection{Correlation between flux and spectral variations}

We also investigated the cross-correlation between the flux-density light
curves and the $F_B/F_R$ curve. In the bottom panel of Fig.~\ref{figure:ccfs} 
we plot the $(F_B+F_R)/2$ vs $F_B/F_R$  and the $F_B$ vs $F_B/F_R$ DCFs (open
circles and solid line, respectively; the $F_R$ vs $F_B/F_R$ DCF is very
similar to the ones plotted in this panel).  Positive lags mean that variations
in the flux-density light curve lead those in the flux ratio curve. 

The two DCFs look very similar. They show a broad hump around zero time lag,
with DCF$_{\rm max}\sim 0.7$ in both cases. This result suggests that the flux
and spectral variations are well correlated. This is not surprising, given the
$F_B/F_R$ vs $(F_B+F_R)/2$ plot shown in the bottom panel of
Fig.~\ref{figure:lcs}. The fact that the maximum DCF value is not larger
than 0.7 can be explained by the significant, albeit of low amplitude, scatter
of the points around the best fitting PL model in this plot. 

The most interesting aspect of both DCFs is the strong asymmetry towards
negative lags.  In order to highlight this effect, in the bottom panel of
Fig.~\ref{figure:ccfs} we also plot the positive-lag $F_B$ vs $F_B/F_R$ DCF
curve reflected about the zero-lag axis (dashed line). It is now evident that
the amplitude of the DCF at lags smaller than $\sim -10$ days is larger than
that at the respective positive lags.  This implies that there are delays
between the flux and spectral variations, with the latter {\it leading} the
former. 

\begin{figure}
   \centering
   \includegraphics[width=8.5cm]{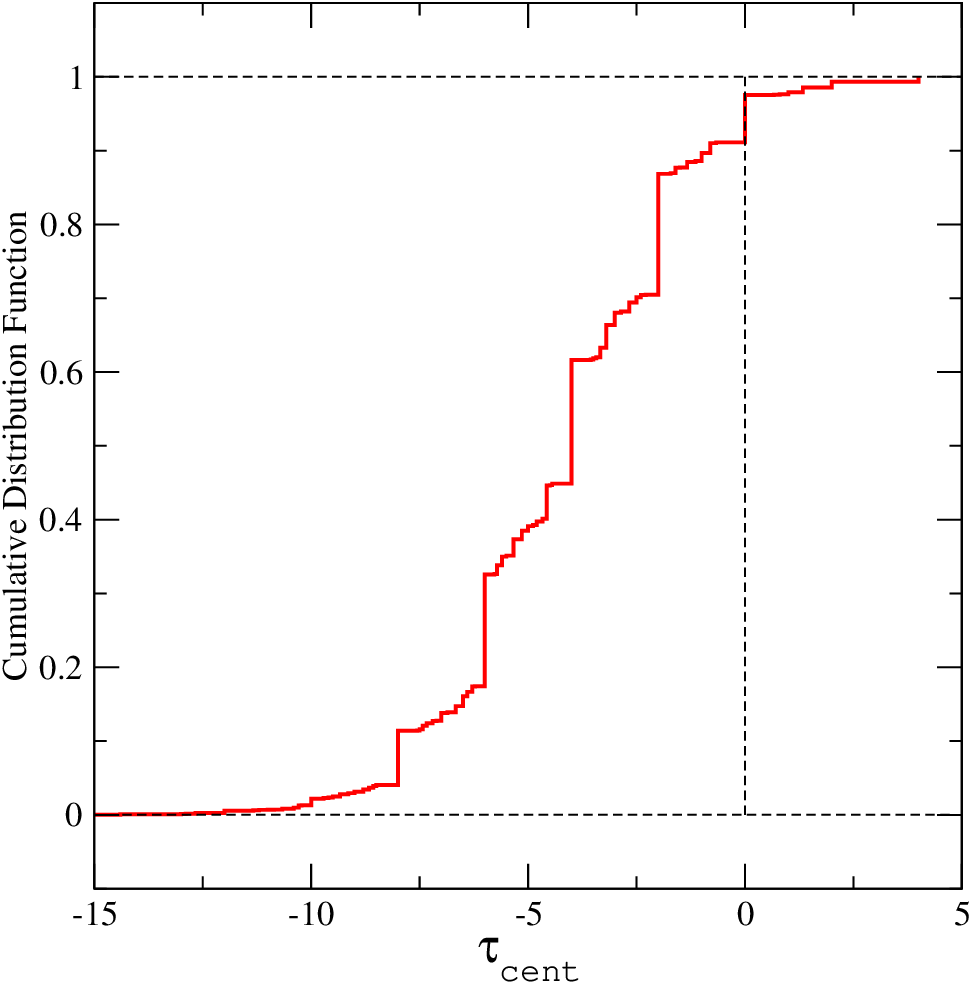}
\caption{The cumulative distribution function of $\tau_{\rm cent}$ in the case
of the $(F_B+F_R)/2$ vs $F_B/F_R$ DCF, as estimated using the results of the
numerical experiment that we describe in Sect.\ 3.2.} 
\label{figure:ccf2} 
\end{figure}

In order to quantify the average delay between the spectral and flux variations,
we estimated the centroid of the $(F_B+F_R)/2$ vs $F_B/F_R$ DCF using the same
method as above. We  found that $\tau_{\rm cent}=-4$ days. In order to
estimate the uncertainty in this measurement, we employed the bootstrap
techniques of \citet{pet98} and created 10000 pairs of simulated light curves.
In Fig.~\ref{figure:ccf2} we plot the sample cumulative distribution of the
centroid value as was estimated using the $\tau_{\rm cent}$ values that we
computed for the 10000 synthetic DCFs of our numerical experiment.

The average centroid value is equal to $-4$ days, and  68.3\% of the synthetic 
DCFs  yield centroid values between $-6.4$ and $-2$ days. Furthermore,  only
2.4\% of all DCFs resulted in $\tau_{\rm cent}$ having a positive value. We
conclude that the delay of $\sim 4$ days that we detect between the  $F_B/F_R$
and the flux variations is significant. 

We should point out that the determination of the DCF centroid is based on a
rather subjective method and depends on the number of the DCF points involved
in the calculation. Furthermore, the DCFs shown in  the bottom panel of
Fig.~\ref{figure:ccfs} show broad humps rather than well defined, narrow peaks.
The broadness of the DCF maxima may imply the presence of more than just a
single delay between the spectral slope and flux variations. 

Having these remarks in mind, we conclude that, although the flux ratio
$F_B/F_R$ follows rather well  the increase/decrease of the source flux (see
bottom panel in  Fig.~\ref{figure:lcs}), the cross-correlation analysis reveals
a subtle detail: $F_B/F_R$ decreases or increases {\it before} the respective
flux changes. This result is significant at the  $97.6$\% level. Our best
estimate of the {\it average} delay between the flux and spectral slope
variations is  $-4^{+2.0}_{-2.4}$ days. 

In order to further investigate this interesting issue, as well as the origin
of the spectral variations, in the following section  we focus our attention on
two best sampled flares in the optical light curves shown in
Fig.~\ref{figure:lcs}. 

\section{Spectral analysis of individual flares}

\subsection{The 1997 flare} 

\begin{figure}
   \centering
   \includegraphics[width=8.5cm]{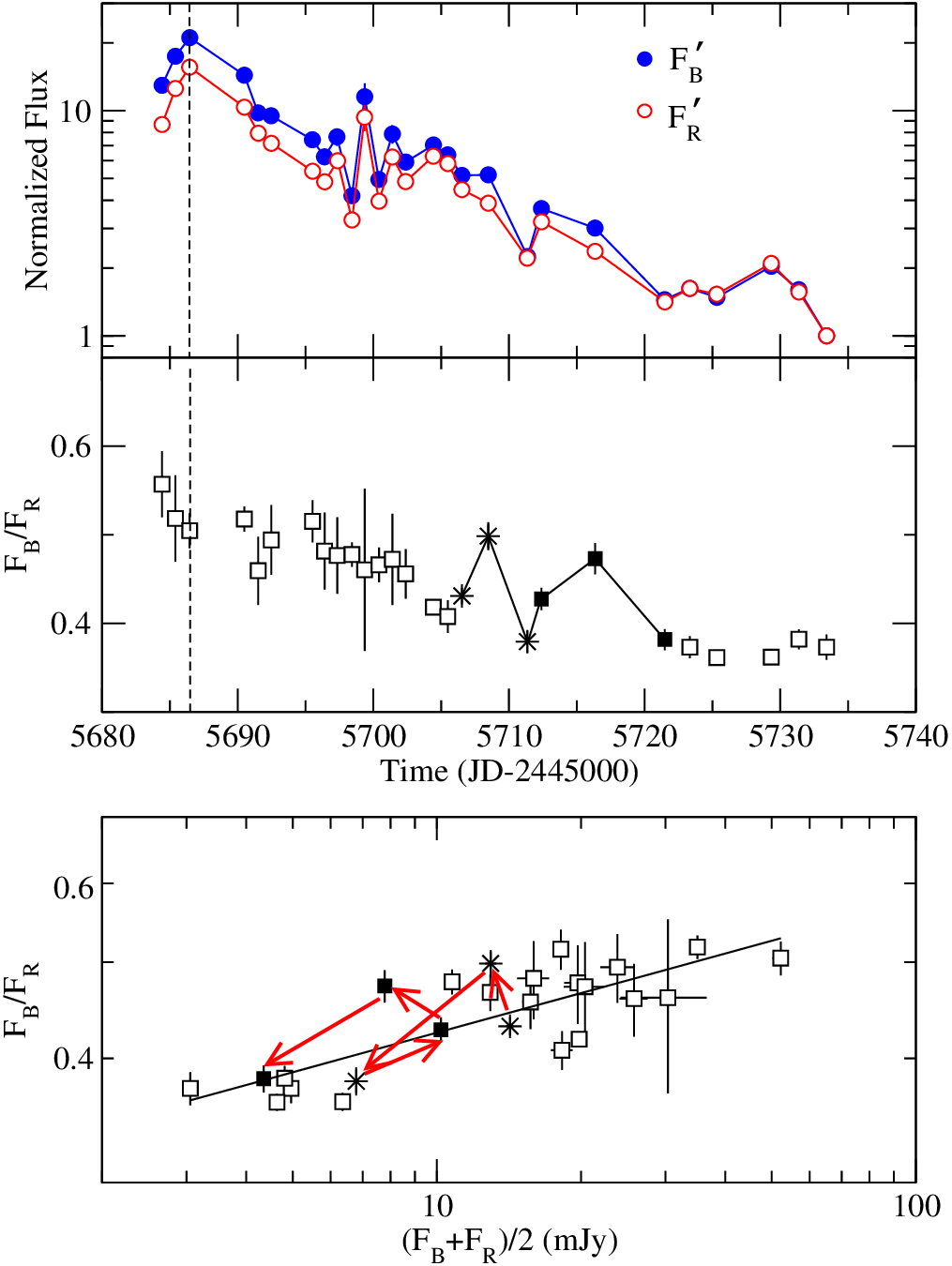}
\caption{Top panel: The last part of the  $B$ and $R$-band light curves of the
1997 flare. The data are normalized to the flux density of the rightmost point.
Middle panel: $F_B/F_R$ plotted as a function of time, using the data shown in
the upper panel. The vertical line indicates the time of maximum
flux. Some points are plotted with asterisks and filled squares for reasons
explained in the text.  Bottom panel: $F_B/F_R$ plotted as a function of
$(F_B+F_R)/2$.   The solid line shows the best-fitting PL model to the data.}
\label{figure:lcp1}
\end{figure}

The 1997 and 2001 large-amplitude optical flares of \bl\ are well sampled in
the light curves that we use in this work (the respective data points in
Fig.~\ref{figure:lcs} are those in the time intervals 5650--5750 and
7050--7150). 

In the top panel of  Fig.~\ref{figure:lcp1} we plot the $B$ and $R$-band light
curves corresponding to the last part of the 1997 flare (filled and open
circles, respectively). The data sample the last few days of the
high-brightness  phase and the subsequent flux decline phase which lasted for
$\sim 50$ days. We do not plot the data before JD = 2450680 because they are
erratic and do not show any other well-defined flux rising or decaying parts.
The light curves are normalized ($F'_B$, $F'_R$)  to their rightmost point
(which, as a result, is set equal to one, for both of them), as this is their
lowest flux point. In this way we can better compare variability amplitudes
between  the two light curves. 

At the flare peak, the $B$ and $R$-band flux densities are $21.1
\pm 0.6$ and $15.6 \pm 0.4$ times higher than the respective  lowest levels. 
The difference between these two values (which represent the maximum
variability amplitude of the light curves shown in Fig.~\ref{figure:lcp1}) is
$5.5 \pm 0.7$. However, as the flux decreases, this difference decreases as
well. For example, $\sim 35$ days after the flare peak, the $B$ and $R$-band
flux densities are only $1.45 \pm 0.04$ and $1.41 \pm 0.02$ times larger than
the respective lowest points. From then on, the $B$ and $R$-band variability
amplitudes are quite similar, indicating that  the flux in the $B$ band 
decreased {\it faster} than the flux in the $R$ band. 

In the middle panel of Fig.~\ref{figure:lcp1} we show the $F_B/F_R$ ratio,
which is clearly variable. This result implies that the optical spectrum of the
source changes, becoming ``softer" (``redder") as the flux decreases.  These
spectral variations are due to the already noticed faster decrease of the
$B$-band flux.  

Interestingly, the data show that $F_B/F_R$ was decreasing even {\it before}
the flux reached its maximum level. The vertical, dashed line in the top panels
of Fig.~\ref{figure:lcp1} indicates the time of the flare peak.   Although the
errors of the two previous  $F_B/F_R$ points are large, the data indicate that
the flux ratio started decreasing in advance of the source flux. This behaviour
is consistent with the ``$(F_B+F_R)/2$ vs $F_B/F_R$" DCF results that we
reported in the previous section. 

The bottom panel of Fig.~\ref{figure:lcp1} shows the ``$F_B/F_R$ vs
$(F_B+F_R)/2$"  plot using the data that are plotted in the top panels of the
same figure. The solid line shows the best-fitting PL model to the data. Its
slope is $\sim 0.13$, almost identical to that of the best-fitting PL model to
the data shown in the bottom panel of Fig.~\ref{figure:lcs}. 
We conclude that a constant difference between the variation rates in the two
bands can produce spectral variations similar to the long-term,
``mildly-chromatic" spectral variations that we observe in \bl\ between 1997
and 2002. 

In the middle and bottom panels of Fig.~\ref{figure:lcp1} we have used different
symbols for some points which appear to deviate significantly from the
generally ``smooth" trend of $F_B/F_R$ decreasing with time. Points plotted with
asterisks and filled squares correspond to the flux  ``humps" that we observe 
in the time intervals 5707--5712 and 5712--5722, respectively. Although
the flux variability amplitude of these ``mini-events" is small, the associated
spectral variations introduce significant scatter around the best-fitting PL
model in the ``$F_B/F_R$ vs $(F_B+F_R)/2$" plot. 

We may thus speculate that some of the scatter around the best-fitting PL model
to the data shown in the bottom panel of Fig.~\ref{figure:lcs} is introduced by
short, low-amplitude events which are accompanied by  spectral variations that
do not follow the general trend of the larger-amplitude flares. 

We note that there exists an interesting correlation between the spectral
variations of these low-amplitude events and the respective flux changes. We
have connected with arrows the points plotted with asterisks and filled squares
in the ``$F_B/F_R$ vs $(F_B+F_R)/2$" plot of Fig.~\ref{figure:lcp1} to indicate
the time evolution  of these ``mini-events" in this plane. We can see that
these points define loop-like structures which evolve in the anti-clockwise
direction. 

\subsection{The 2001 flare}

\begin{figure}
   \centering
   \includegraphics[width=8.5cm]{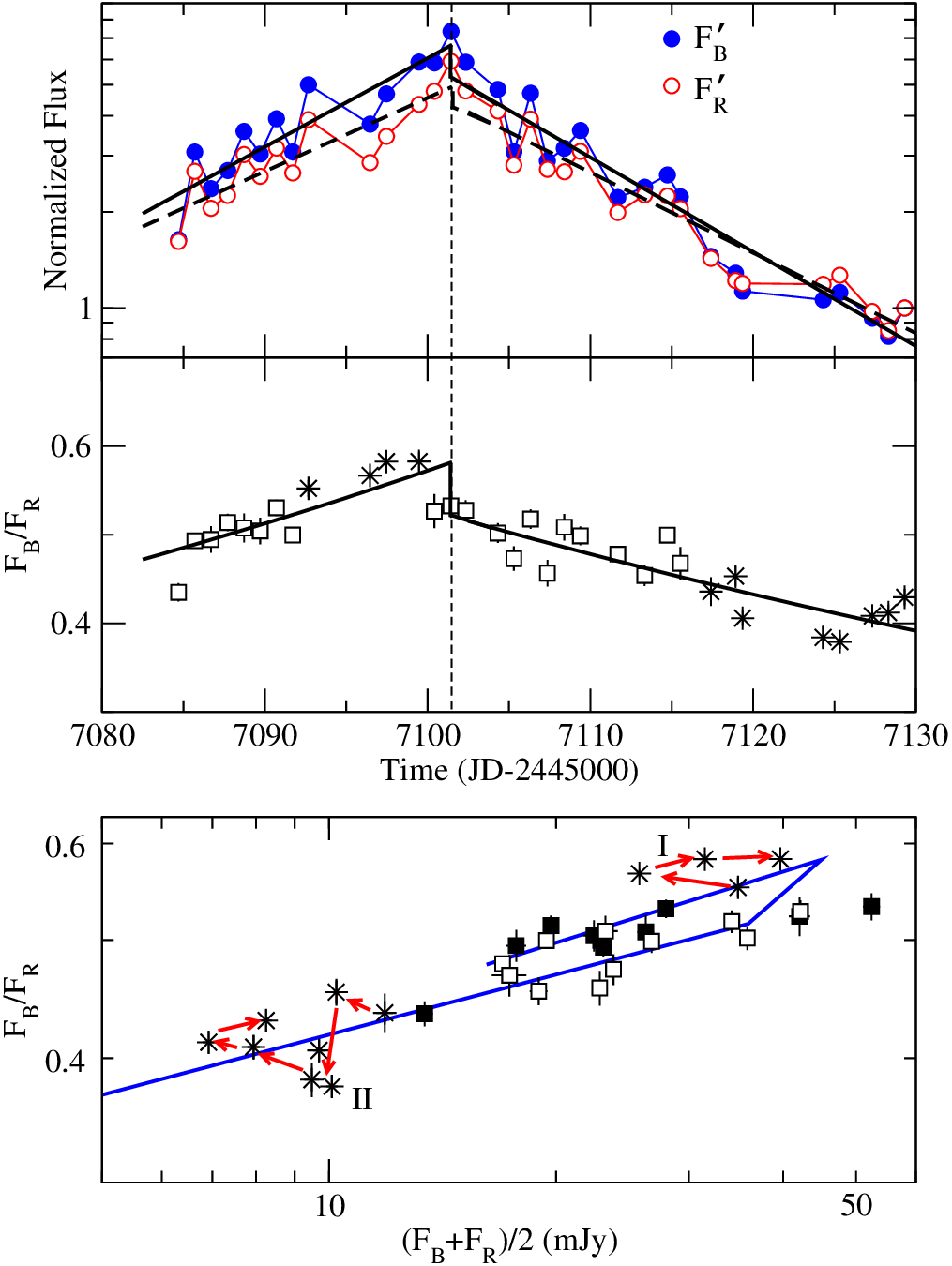}
\caption{Same as Fig.~\ref{figure:lcp1}, but for the 2001 flare.  The solid and
dashed lines in the top panel show the best-fitting exponential functions to
the $B$ and $R$-band data, respectively. The solid line in the middle panel
shows the resulting $F_B/F_R$ ratio. Filled and open squares in the bottom
panel show the data during the flux rising and decaying phases, respectively.
Some points in the middle and bottom panels are  plotted with asterisks for
reasons explained in the text.}
\label{figure:lcp2}
\end{figure}

During the BL Lac 2001 WEBT campaign, two large-amplitude flares, the first in
mid 2001 and the second in  early 2002, were detected \citep{vil04a}. The 2001
flare was double-peaked. The second peak is well sampled in the WEBT light
curves that we use in this work. The respective data are plotted in the top
panel of Fig.~\ref{figure:lcp2}. Just like in Fig.~\ref{figure:lcp1}, we have
normalized both light curves to the flux density of their rightmost point. In
the middle panel we show $F_B/F_R$ as a function of time, and in the bottom
panel the same quantity is plotted as a function of $(F_B+F_R)/2$.  The
advantage of the 2001 light curves is that they sample well both the flux
rising and flux decaying  parts of the flare. 

We fitted the rising and decaying parts of the light curves with an
exponentially increasing, $F(t)\propto \exp[(t-t_{0})/\tau_{\rm r}]$, and
decreasing, $F(t)\propto \exp[-(t-t_{0})/\tau_{\rm d}]$,  function,
respectively ($t_{0}$ represents the time of the flare peak). The best-fitting
models are also plotted (normalized as the light curves) in the top panel of 
Fig.~\ref{figure:lcp2} (solid and dashed line for the $B$ and $R$-band light
curves, respectively). 

The model describes rather well the overall flux evolution, during both the
rising and decreasing phases of the flare. Our best-fitting $\tau$ values are:
$\tau_{{\rm r},B}=16, \tau_{{\rm r},R}=18, \tau_{{\rm d},B}=13$,  and
$\tau_{{\rm d},R}=15$ days  (with an error of $\pm 0.8$ days in all cases). 
The differences among these best-fitting  $\tau$ values suggest that:  a) the
flare is not symmetric, as the flux decrease is faster than the flux increase,
i.e.\ $\tau_{{\rm d}}$ is smaller than $\tau_{{\rm r}}$ in both bands (by the
same amount of $\sim 3$ days), and b) the flare evolves faster in the $B$ than
in the $R$ band, as both the rise and decay $\tau_{B}$ values are smaller than
the respective $\tau_{R}$ values by about 2 days.

The second result can explain the observed spectral variations. The solid line
in the middle panel of Fig.~\ref{figure:lcp2} shows the $F_B/F_R$ curve derived
from the best-fitting $B$ and $R$-band exponential functions. Obviously, the
$F_B/F_R$ model curve fits the data well. At the beginning of the flare, the
spectrum hardens as the flux increases, because the $B$-band flux increases
faster than the flux in the $R$ band ($\tau_{{\rm r},R}>\tau_{{\rm r},B}$).
After the flare peak,  the spectrum softens, because the  $B$-band flux
decreases faster than the $R$-band flux. 

The vertical dashed line in the upper two panels of Fig.~\ref{figure:lcp2} 
indicates the time of the flare peak. One can see that the change of the
spectral evolution from spectral hardening (i.e. $F_B/F_R$ increasing) to
spectral softening ($F_B/F_R$ decreasing) happens a couple of days {\it before}
the flux reaches its maximum. This is consistent with the DCF results that we
report in Sect.\ 3.2. Furthermore, the data points in the ``spectral slope vs
time" plot indicate that the change in the spectral evolution is rather abrupt.
This is also clearly indicated by the significant ``break'' that appears in
the  $F_B/F_R$ model curve near the flux peak. 

In the bottom panel of Fig.~\ref{figure:lcp2} we show the ``$F_B/F_R$ vs
$(F_B+F_R)/2$" plot. Filled and open squares correspond to the data during the
rising and decaying phases, respectively. A PL model of slope $\sim 0.13$ (not
plotted for clarity reasons), equal to that of the 1997 flare, describes well
the  ``$F_B/F_R$ vs $(F_B+F_R)/2$" relation. 

The solid (blue) lines in the same panel show the  ``spectral-slope vs
brightness" relation when we consider the best-fitting exponential rising and
decaying functions, which are  plotted in the upper panel of
Fig.~\ref{figure:lcp2}. In agreement with the data points  (open squares lay
systematically below the points plotted with filled squares), and because of
the abrupt change of the spectral slope near the flare maximum,  the
``spectral-slope vs brightness"  model curve during the flux decay phase lies
below the respective flux rise model curve.

Just like during the 1997 flare, low-amplitude  flux variations introduce
significant scatter in the ``$F_B/F_R$ vs $(F_B+F_R)/2$" plot. The points
plotted with asterisks in the middle and bottom panels of 
Fig.~\ref{figure:lcp2} highlight the spectral variability behaviour of the
source during such ``events". 

For example, during the time interval $\sim 7093$--7102 the flux decreased and
then increased until the flare reached its peak.  The respective points in the
``spectral-slope vs flux" plot follow a well defined loop, which evolves
clockwise (loop ``I"), as the arrows clearly show in the bottom panel of
Fig.~\ref{figure:lcp2}. 

We observe loop-like patterns during the flux decaying part of the flare as
well.  The clearest example is the spectral evolution during the end of the
flare, i.e.\ after $\sim 7117$.  
can see that these points correspond 
smooth flux increase, which becomes more 
In the $F_B/F_R$ vs $(F_B+F_R)/2$ plot, the respective data, plotted with
asterisks, follow a clockwise evolving loop-like pattern (loop ``II"). Note
that $F_B/F_R$ has already started increasing a few days before the flux
increase. 

\section{Discussion and conclusions}

\bl\ is one of the most frequently observed blazars, at all frequencies.  It
has been observed for more than a century in the optical band. The observations
in the last years have been very intense, due to the recent WEBT campaigns to a
large extent. The light curves that have resulted from these observations are
densely sampled over a long time period ($\sim 8$ years). Although  optical
observations of the source will be carried out in the  future as well, we
believe that the current light curves are of sufficient quality (in terms of
length and sampling) to allow the investigation of its``typical" variability
characteristics. 

\citet{vil04a} already showed  that the variations in the optical light curves
of \bl\ can be interpreted in terms of a ``strongly" and  a ``mildly-chromatic"
component, which operate on short (i.e. $<1$ day)  and longer time scales,
respectively.  They also showed that the short and longer time scale 
components determine a ``bluer-when-brighter" slope of $\sim 0.4$ and $\sim
0.1$, respectively, in a ``$B-R$ vs $R$" plot. 

In this work, we continued the investigation of \citet{vil04a}, and we studied 
in more detail the long-term, mildly-chromatic spectral variations that these
authors had already noticed. Our main results are as follows: 

a) There is an indication that  the $R$-band variations on time scales longer than a day are slightly delayed with respect to
the variations in the $B$ band. However, this is a low amplitude effect and 
longer light curves are needed to confirm it. In any case, even if real, this
effect  cannot explain the significant, long-term, optical   spectral
variations that we observe. 

b) A constant difference between the rates with which the flux rises/decays in
the $B$ and $R$-band light curves {\it can} explain the spectral variations
that we observe in \bl. 

For example, we find that the 2001 flare data are  consistent with
exponentially rising/decaying functions. When the flux rises, 
$F_{B}/F_{R}\propto {\rm exp}[(t-t_{0})(\tau_{{\rm r},R}-\tau_{{\rm r},B})/
(\tau _{{\rm r},R} \tau_{{\rm r},B}) ]$. Using our best-fitting  $\tau_{{\rm
r},R}$  and $\tau_{{\rm r},B}$ values, we find that $F_{B}/F_{R}\propto
F_{R}^{0.12}$.  In the ``$B-R$ vs $R$" plane this relation corresponds to a
line of slope $\sim 0.12$, consistent with the findings of \citet{vil04a}. We
get a similar  result when we consider the ratio $F_{B}/F_{R}$ during the flux
decay phase. In this case,  $F_{B}/F_{R}\propto F_{R}^{0.15}$. 

Obviously, the rise and decay time scales, as well as the maximum variability
amplitude, are not identical in  all the ``flare-like" events in \bl.  In 
Fig.~\ref{figure:lcs} one can detect  many low-amplitude events, which last
much less than the large-amplitude 2001 flare. What our results suggest is
that  the characteristic time scales of the flares  cannot change {\it
arbitrarily} from one to the other. As an  example, if we assume exponentially
rising/decaying flares,  their time scales $\tau_{{\rm r/d},R}$ and 
$\tau_{{\rm r/d},B}$ should change in such a way so that the ratio  
$(\tau_{{\rm r/d},R}-\tau_{{\rm r/d},B})/ \tau_{{\rm r/d},B}$  always remains
roughly equal to $\sim 0.1$. In this way, the relation between  the spectral
slope and flux variations can be explained. 

c) We find that the flux variations {\it follow} the spectral variations by an
average delay of $\sim 4$ days. This is the first time that such an effect has
been observed in \bl. A similar behaviour, in the optical band, has also been
observed recently in 3C 66A \citep{boe05}. This effect can be clearly seen in
the well-sampled 1997 and 2001 flares. The DCF results we report in Sect.\ 3.2
suggest that  this delay operates most of the times in \bl. 

If the  ``spectral-slope vs brightness" plot can be explained by a constant
difference  between the variability  time scales in the various bands (as we
showed above),  then the spectral hysteresis effect can be explained if one
assumes that the variability time scale does not remain constant during the
flare flux rise phase. The spectrum can start softening before the source
reaches its maximum flux if the variability time scale, which determines the
flux evolution, starts increasing before the peak at a rate which increases
with frequency, or vice versa. 

d) Although a power-law model describes well the  ``bluer when brighter" trend
in the ``spectral-slope vs brightness" plot, it cannot provide a statistically
acceptable fit to it. There exists significant scatter around the best-fitting
model line. 

To some extent, this scatter can be explained as the ``remnant" of the fast,  
strongly-chromatic flares of \citet{vil04a}  in the 1-day binned light curves
we use. It is quite possible that in many cases the  daily average points do
not correspond to the ``mean" of the fast flares  but rather to  a peak or a
dip of the fast flares. 

At the same time though, as our analysis in Sect.\ 4 shows, some of the scatter
is also due to the fact that, occasionally,  the spectral evolution of the
source defines loop-like  structures in the ``spectral-slope vs brightness"
plot. In fact, if the ratio $F_B/F_R$ decreases/increases a few days before
$F_B$ and $F_R$ (see result ``c" above), then most flares that last for more
than 5--10 days should show clockwise evolving loop structures in the 
$F_B/F_R$ vs flux plot. This is what we observe during the 2001 flare.

Anti-clockwise evolving loops (like the ones observed in the flux decaying
phase of the 1997 flare) imply that the spectral slope variations {\it follow}
those in the flux light curves. The spectral-slope vs flux DCF results suggest
that these must be rare events. In any case it will be interesting to
investigate this issue in the future, using the even longer light curves that
will be available then. 

In conclusion, our two main findings are: i) the confirmation of the 
bluer-when-brighter  mild chromatism of the BL Lac long-term optical
variations, which is not caused by any delays between the $B$ and $R$-band
light curves, but can be explained if the flux rise/decay rates in the two
bands  are different, by a fixed amount, and ii) the discovery that the
spectral-slope variations tend to lead the flux variations, by a few days.

\citet{vil04a} explained the bluer-when-brighter behaviour in terms of a
Doppler factor, $\delta$, variation on a convex spectrum. Indeed, an increase
in $\delta$ implies both an increasing flux ($F_\nu \sim \delta^3$; see e.g.\
\citealt{vil99}) and a change in the observed frequency ($\nu \sim \delta$).
The former effect causes the flux to increase at all frequencies, thus the
appearance of a flare, while the intrinsic spectrum of the source is also
``boosted" towards higher frequencies. If this spectrum is convex, we will
observe a harder and harder part of it in our optical band, during the rising
part of the flare, and vice versa during the dimming phase, i.e.\ the
bluer-when-brighter trend. In other words, a change of $\delta$ on a convex
spectrum makes flux variations in $B$ and $R$ bands happen simultaneously, but
the $B$-band flux rises/decays faster than the $R$-band one.

Moreover, if the $\delta$ variation is due to a change of the jet viewing angle
($\delta=[\gamma(1-\beta \cos \theta)]^{-1}$) and the jet is inhomogeneous and
curved \citep[see e.g.][]{vil07}, we will also have a tendency to a progressive
reddening of the spectrum during the flare. Indeed, since the jet is
inhomogeneous, it is composed of ``slices" emitting different frequency ranges,
which are redder and redder going outwards, along the jet curvature. Thus, as
the jet approaches the line of sight, the $\delta$ variation will affect first
the bluer and then the redder parts, implying a ``reddening" of the spectrum.
However, this reddening is opposed by the ``bluer-when-brighter" trend during
the  flare rising phase. The combination of the two effects can result in a
softening of the spectrum before the flare peak, as observed. In other words,
the concomitance of the  ``bluer-when-brighter" trend with the progressive
reddening, both of geometrical-kinematic origin, can (at least qualitatively)
account for the changes in the spectral slope we analyzed in this paper.

A further observing as well as modeling effort is needed to clarify and
quantify this interpretation. On the observational side, apart from \bl,
the WEBT archive already stores long and well sampled light curves for three
more sources, and more data will appear in the near future. The  analayis of
these light curves in a way similar to what we presented in this work, and the
comparison of the results with those presented here, will hopefully help us
understand better the variability mechanism in BL Lacs.

\begin{acknowledgements}
This work was partly supported by the European Community's Human Potential Programme
under contract HPRN-CT-2002-00321 (ENIGMA).
\end{acknowledgements}

\bibliographystyle{aa}

\end{document}